# Heterogeneous Memristive Devices Enabled by Magnetic Tunnel Junction Nanopillars Surrounded by Resistive Silicon Switches

Yu Zhang, Xiaoyang Lin, Jean-Paul Adam, Guillaume Agnus, Wang Kang, Wenlong Cai, Jean-Rene Coudevylle, Nathalie Isac, Jianlei Yang, Huaiwen Yang, Kaihua Cao, Hushan Cui, Deming Zhang, Youguang Zhang, Chao Zhao, Weisheng Zhao*, and Dafine Ravelosona

Mr. Y. Zhang, Dr. X. Lin, Dr. W. Kang, Mr. W. Cai, Dr. J. Yang, Dr. H. Yang, Mr. K. Cao, Mr. H. Cui, Mr. D. Zhang, Prof. YG. Zhang, Prof. C. Zhao, Prof. W. Zhao
Fert Beijing Institute, BDBC Beihang University, Beijing 100191, China
E-mail: weisheng.zhao@buaa.edu.cn
Mr. Y. Zhang, Dr. J. P. Adam, Dr. G. Agnus, Dr. J. R. Coudevylle, Mrs. N. Isac, Dr. D. Ravelosona
Centre de Nanosciences et de Nanotechnologies, University of Paris-Sud, Université Paris-Saclay, Orsay 91405, France
Mr. K. Cao, Mr. H. Cui, Prof. C. Zhao
Institute of Microelectronics, Chinese Academy of Sciences, Beijing 100029, China

Emerging non-volatile memories (NVMs) have currently attracted great interest for their potential applications in advanced low-power information storage and processing technologies. Conventional NVMs, such as magnetic random access memory (MRAM) and resistive random access memory (RRAM) suffer from limitations of low tunnel magnetoresistance (TMR), low access speed or finite endurance. NVMs with synergetic advantages are still highly desired for future computer architectures. Here, we report a heterogeneous memristive device composed of a magnetic tunnel junction (MTJ) nanopillar surrounded by resistive silicon switches, named resistively enhanced MTJ (Re-MTJ), that may be utilized for novel memristive memories, enabling new functionalities that are inaccessible for conventional NVMs. The Re-MTJ device features a high ON/OFF ratio of >1000% and multilevel resistance behaviour by combining magnetic switching together with resistive switching mechanisms. The magnetic switching originates from the MTJ, while the resistive switching is induced by a point-switching filament process that is related to the mobile oxygen ions. Microscopic evidence of silicon aggregated as nanocrystals along the edges of the nanopillars verifies the synergetic mechanism of the heterogeneous memristive device. This device may provide new possibilities for advanced memristive memory and computing architectures, e.g., in-memory computing and neuromorphics.

## 1. Introduction

Non-volatile memories (NVMs) combined with novel computing architectures have recently been considered as the most promising solution to overcome the "memory wall" of von-Neumann computing systems[1-3]. For instance, in-memory computing architectures based on closely integrating fast NVMs with logic functions have been proposed to minimize the power consumption and pave the way towards normally-off/instant-on computing[4,5]. Meanwhile, neuromorphic computing inspired by the human brain exploits the resistive features of NVMs as artificial synapses and neurons and has already triggered a revolution for non-von-Neumann architectures[6-8]. Along this direction, two of the most promising NVMs, *i.e.*, magnetic random access memory (MRAM)[3,9-11] and resistive random access memory (RRAM)[3,12-14], have attracted increasing interest. Tremendous efforts have been involved and amazing advances have been achieved in this field. Nevertheless, a few issues still exist and should be addressed before popular applications can be achieved. For example, multilevel resistance states can be achieved in either magnetic tunnel junctions (MTJs, the core device of MRAM) or RRAM devices. For MTJ devices, there are two main methods to obtain the multilevel resistance states, either by taking advantage of the stochastic behaviour of the magnetic switching[15], or by using vertical stacked MTJs as multilevel cell[16]. However, both of methods suffer from a challenge of a relative low tunnel magnetoresistance (TMR) ratio (<250% to date), which is a key limitation for high density and high reliability applications[17-19]. Regarding RRAM devices with high ON/OFF ratios, they can provide the perfect multilevel resistances required for applications, but its relative low access speed and endurance issues are limited by its intrinsic mechanism as electrochemical reduction and Joule heating process, which have become an intrinsic drawback for computing tasks.

Therefore, an NVM that eliminates these shortcomings, for example, by integrating MRAM and RRAM into a single device, is still highly desired. Several recent studies based on an MgO-based MTJ have exhibited both magnetic switching (MS) and resistive switching (RS), enabling such possibilities[20-22]. However, those devices suffer from a trade-off between MS and RS as the MgO layer acts as both a tunnel barrier for MS and an insulator for RS. In this work, we report a novel heterogeneous memristive device composed of an MTJ nanopillar surrounded by resistive silicon switches, named a resistively enhanced MTJ (Re-MTJ), that can combine the advantages of both RRAM and MRAM for realizing advanced memristive memories. Different from previous studies, the RS in our proposed Re-MTJ originates from a point-switching filament process related to the mobile oxygen ions [23-26], while the MS from the MTJ. The MS and RS in the Re-MTJ can be individually optimized. Furthermore, our tests show that the Re-MTJ device can achieve a rather high ON/OFF ratio of >1000% and multilevel resistance behaviour; both of these characteristics are urgently required in advanced NVMs. In addition, the Re-MTJ may provide new functionalities that are inaccessible to conventional NVMs, e.g., for in-memory computing and neuromorphic computing as non-von Neumann computing architectures.

## 2. Results and Discussion

**Figure 1a** shows a schematic view of the Re-MTJ device made of an in-plane magnetized CoFe(B)-MgO MTJ nanopillar and a $SiO_x$-based polymer encapsulation layer. In the MTJ structure, the MgO tunnel barrier is sandwiched by two CoFe(B)-based magnetic thin films corresponding to the free and reference layers. Driven by either an in-plane magnetic field or a spin-polarized current, the MTJ can be switched between parallel (P) and anti-parallel (AP) states, leading to two resistance states, namely, $R_P$ and $R_{AP}$, respectively. Figure 1b shows a cross-sectional transmission electron microscope (TEM) image of the device. The Re-MTJ devices were fabricated in a method similar to the conventional MTJs. The magnetic films were patterned into submicron-sized ellipses with dimension ranging from $80 \times 160$ nm$^2$ to $100 \times 240$ nm$^2$ using electron beam lithography (EBL) and an angle-optimized ion beam etching (IBE) process. Then, the MTJ nanopillars were encapsulated by a Si-O-based insulator fabricated by baking a polymer (Accuflo)[27] at a reduced temperature. As shown in Figure 1c, the cross-sectional TEM image of MTJ stacks after annealing indicates the presence of the crystallized MgO tunnel barrier and CoFe(B) layers, which has been reported as NaCl-structure and body-centred cubic (bcc) structures respectively[9,28].

**Figure 2** indicates the very peculiar resistive behaviour of the device under an applied voltage resulting from the combination of current-induced magnetization switching (CIMS) and voltage-induced resistive switching (VIRS). Indeed, as shown in the typical I-V curves of Figure 2a, bipolar (in which positive and negative voltages have opposite effects) VIRS was observed in the absence of external magnetic fields, with a maximum voltage of 0.8 V. For this measurement, the MTJ was set to either the P or AP state using the external magnetic field before the measurement. Clear resistive switching behaviour was observed from which the SET and RESET voltages could be identified. Figure 2b shows the corresponding R-V curve for the P and AP configurations, in which the low-resistance state (LRS) and high-resistance state (HRS) can be observed. Here, the LRS is approximately $600 \Omega$, irrespective of the magnetic configurations, and the HRS is approximately $1100 \Omega$ and $1300 \Omega$ for the P and AP states, respectively. We note that the low TMR (around 20%) presented here is related to a degraded crystalline structure of MgO layer with only 0.8 nm, for the facts that MgO(001) acts as a template to crystallize CoFe(B) and a bcc structure of CoFeB is crucial for gaining high TMR[9]. Since the MgO layer grown on an amorphous CoFeB is only expected to have highly oriented polycrystalline MgO(001) structure when $t_{MgO} < 5ML$ (1.05 nm)[29], an optimization for TMR is possible for increasing the thickness of MgO layer. In this example, the typical ON/OFF ratio for the AP state was approximately 120%, but ratios reaching 1000% were observed (see **Figure S2** in the Supporting Information). Notably, magnetization switching between the P and AP states was not observed during the voltage sweep. Additionally, we observed that the SET and RESET voltages were independent of the magnetic configuration. Figure 2c shows the R-V curve under the in-plane magnetic fields[30] that were used to assist CIMS. Pure CIMS was not obtained here at low voltages due to the thickness of the free layer. The maximum voltage applied was below 0.2 V in order to avoid VIRS and maintain the HRS. We observed CIMS assisted by the magnetic fields of $H_{ext}^{AP \to P} = +110$ Oe and $H_{ext}^{P \to AP} = -104$ Oe with typical current densities of $J_c^{AP \to P} = 1.7 \times 10^5\,\text{A cm}^{-2}$ and $J_c^{P \to AP} = 0.8 \times 10^5\,\text{A cm}^{-2}$, respectively. These results show that CIMS and VIRS could be controlled independently. To observe both effects, a voltage was applied between $\pm 0.8$ V under a magnetic field (Figure 2d). In this case, we clearly observed that the VIRS and CIMS effects could act simultaneously.

The results shown in Figure 2 are different from those of previous findings involving RS due to the filamentary current path in the MgO barrier. First, in our experiment, we observed bipolar switching instead of unipolar switching (SET and RESET were caused by applying voltages with the same polarity). In addition, we observed both MS and RS in the same R-V loop; this result suggests two independent origins for the CIMS and RS processes. To gain more insight, we carefully investigated the microstructures of the elements. The nanofabrication process of the device consisted of encapsulating the CoFe(B)-MgO nanopillars with a $SiO_x$ insulator in contact with the edges of the nanostructure (see **Figure 1b and 3a**). In the following, we provide evidence that the VIRS behaviour was induced by the presence of the resistive Si filaments at the edges of the nanopillars. Microscopic structure characterizations were performed using energy-dispersive X-ray spectroscopy (EDS). Figure 3b and 3c indicates the presence of the Ta and Si elements detected by measuring the characteristic peaks of the $Ta\text{-}L_\alpha$ (8.145 KeV) and $Si\text{-}K$ series lines, respectively. Note that since $Ta\text{-}M_\alpha$ (1.709 KeV) and $Si\text{-}K_\alpha$ (1.739 KeV) were separated by only 30 eV, the detector could not resolve these lines[31,32]. The detection results of both the Si and Ta elements overlapped, as seen in Figure 3c. Thus, the comparison between Figure 3b and 3c clearly evidenced that Si aggregation occurred along the sidewall of the nanopillars with a typical width of 5-10 nm. In addition, the high-resolution TEM (HRTEM) images of the nanodevice indicate the presence of nanocrystals with a typical size of 5-10 nm embedded in the amorphous $SiO_x$ along the edges of the nanopillars (see Figure 3d). The microstructural analysis and the electrical results of Figure 2 are consistent with the results from recent studies, which indicated that RS in a $SiO_x$ matrix can be induced in the presence of embedded Si nanocrystals[23,33-35]. More precisely, when an SET voltage is applied, the Si nanocrystals can grow locally by favouring an electrochemical reduction process from $SiO_x \rightarrow Si$. This process induces a Si pathway (Si filaments) along the current flow direction, whereas a RESET voltage can favour the $Si \rightarrow SiO_x$ inverse process. This mechanism corresponds to a point-switching filament process involving local breakage and bridge evolution.

One important question is related to the presence of Si nanocrystals in our devices. It has been shown that the forming process of Si nanocrystals can be induced within pure $SiO_x$ matrixes at low temperatures by etching the $SiO_x$[23,36]. In this case, the Si filaments can germinate at the edges of the $SiO_x$ elements due to the presence of defects. In our case, the $SiO_x$ matrix surrounding the nanopillars was obtained by spinning a polymer (Accuflo) and transforming it into an insulator using an annealing process at approximately 300 °C. During the annealing process, the edges of the nanopillars involving damage induced by the etching process (see Experimental Section and Supporting Information Note 6) could serve as seed interfaces to nucleate the Si nanocrystals. In addition, the crystalline character of the MTJ may have also favoured the germination of the Si nanocrystals. Indeed, an EDS linescan measured from the SiO$_x$ matrix into the MgO barrier (see Figure 3e and 3f) indicated that both Si and O aggregated at the edges of the nanopillars on a scale of 10 nm with a ratio of silicon to oxygen elements that was much higher at the edges than in the SiO$_x$ matrix. The fact that bipolar behaviour was observed here for the SET and RESET processes may have been related to the presence of mobile oxygen ions.

Based on the analyses described above, a proposed schematic of the Re-MTJ device is presented in **Figure 4a** that consists of an MTJ-based element connected in parallel with a Si filament element. Such a device structure indicates that four distinct configurations with different resistance states can be achieved (see Figure 4b); this result is in agreement with the experimental results (see Figure 2d). When the Si filaments are not conductive (RESET process), the current mainly goes through the MTJ, resulting in an HRS, and when the Si filaments becomes conductive (SET process), the current mainly flows through the Si filaments, resulting in an LRS. To further verify the proposed device model, simulations were performed using a compact model that integrated a physical-based STT-MTJ[37] and a bipolar metal-insulator-metal (MIM) resistive junction[38] connected in parallel (see Figure 4c). Using the parameters of $R_{AP} = 1390\ \Omega$, $R_P = 1160\ \Omega$, $R_{HRS} = 64500\ \Omega$, and $R_{LRS} = 660\ \Omega$, the R-V curve of Figure 2d that combines the features of CIMS and VIRS could be well reproduced. Furthermore, another interesting feature related to the microstructural properties of the devices is the strong correlation between the ON/OFF ratio of the RS and the TMR value of the MS (see Figure 4d). In particular, the ON/OFF ratio increased when the TMR value was reduced. This result suggests that when the TMR ratio was low, a point-switching filament process could occur, whereas when the TMR was higher, the conduction through the Si filaments was not active. Notably, the ON/OFF ratio and TMR behaviour in these devices originated from the formation of an Si pathway in the $SiO_x$ matrix[23] and the $\Delta_1$ Bloch states filtering at the CoFe(B)/MgO interface[39], respectively. The oxygen ion movement from the $SiO_x$ matrix towards the MTJ nanopillars promoted the nucleation of the Si nanocrystals and affected the Fe-O bonds at the CoFe(B)/MgO interface[28,39-41]; these processes resulted in a high ON/OFF ratio but a low TMR. As a result, the mobile oxygen ions near the edges of the nanopillars (see Figure 4a) played a joint role in both the ON/OFF ratio and the TMR value.

The multilevel states of the Re-MTJ device were investigated and are presented in **Figure 5**. Figure 5a shows seven consecutive R-V curves indicating that different resistance states could be reached using a single device. Each R-V curve corresponds to a different degree of the Si oxidation pathway, and the pathways were randomly induced by the combination of a local strong electric field and heating during the point-switching filament process. Figure 5b presents a Re-MTJ device that exhibited eight different states by combining two magnetic states (P and AP) with four different resistance states of the Si filaments. A larger TMR ratio was achieved for higher resistance states; this result reflects that the lower resistance of the MTJ dominated the current pathway. Furthermore, the data retention of the Re-MTJ device was tested for four different resistance states (see Figure 5c). All the configurations exhibited robust non-volatile properties. Although the resistances between LRS with P and LRS with AP are close (see Figure S5 in the Supporting Information), it can be improved by enhancing the TMR ratio with the further optimization of fabrication process.

## 3. Conclusion

In conclusion, we fabricated a heterogeneous memristive device, Re-MTJ, an advanced NVM device that combines the merits of MRAM and RRAM. The Re-MTJ device integrates an MTJ nanopillar encapsulated by a $SiO_x$-based polymer with surrounding resistive silicon filaments. We observed both MS and RS in the Re-MTJ; these behaviours originated from the MTJ and silicon filaments, respectively. We reported a rather high ON/OFF ratio and multilevel resistance behaviour owing to the point-switching silicon filament process. These properties are rather preferable for high-reliability and high-density memory applications. The presence of nanocrystals within the silicon aggregates was confirmed by the microscopic structure characterizations. The proof-of-concept demonstration here involved low TMR values, but in principle, our approach can enable one to independently optimize the properties of the MTJ and silicon elements. The Re-MTJ device, with the merits of a high ON/OFF ratio, long endurance and multilevel resistance behaviour, can certainly benefit the advancement of memristive memory and computing architectures, such as in-memory computing and neuromorphics.

## 4. Experimental Section

*Sample preparation:* The magnetic multilayers were deposited onto $SiO_2$-coated Si wafers using a combination of RF and DC sputtering in a Canon-Anelva system. From the substrate side, the MTJ structure consisted of the following layers: Ta(5)/Ru(15)/Ta(5)/Ru(15)/Ta(5)/Ru(5)/PtMn(20)/CoFeB(1.5)/CoFe(2.0)/Ru(0.85)/CoFeB(1.5)/CoFe(1.5)/MgO(0.8)/CoFe(1.5)/CoFeB(1.5)/Ru(2)/Ta(5)/Ru(10) (the numbers are the nominal thicknesses in nanometers). The bottom and top layers, Ta(5 nm)/Ru(15 nm)/Ta(5 nm)/Ru(15 nm)/Ta(5 nm)/Ru(5 nm) and Ru(2 nm)/Ta(5 nm)/Ru(10 nm), respectively, were designed for the CIPT measurements using a CAPRES microprobe tool. The typical TMR ratio and the resistive-area product of the unpatterned films were ~ 144% and ~ 19 $\Omega\ \mu m^2$, respectively. Then, the annealing was performed at 350 °C for 1 h with an in-plane applied magnetic field of 1 T under a vacuum of $10^{-6}$ Torr. After the deposition, the multilayers were patterned into submicron-sized ellipses by electron beam lithography and ion beam etching. A low-temperature curing process of Accuflo was utilized for encapsulating the patterned structure.

*Transport measurements:* The fabricated devices were characterized using dc-transport measurements under in-plane magnetic fields (with a precision below $1\times10^{-3}$ Oe) with a two-probe geometry at room temperature. A bias voltage (or current) was applied to the top electrode, while the bottom electrode was grounded. The voltage-pulse (or current-pulse) durations were $\tau_p = 200$ ms, and the remanent resistance of the Re-MTJ device was measured under a low bias between each voltage (or current) change.

*Model simulation:* The compact model of the device was written using Verilog-A language and evaluated in a Cadence Spectre environment. The compact model integrated a physical-based STT-MTJ and a bipolar MIM resistive junction connected in parallel. Magnetic fields were not considered in the simulation.

**Acknowledgements**

The authors thank Mr. Li Huang, Prof. Xiufeng Han, Mr. Sylvain Eimer and Dr. Fabien Bayle for their technical support as well as Dr. Qunwen Leng for fruitful discussions. The authors acknowledge the financial support by the Chinese Scholarship Council (CSC), the projects from the National Natural Science Foundation of China (No. 61571023, 61501013, 51602013 and 61627813), Beijing Municipal of Science and Technology (No. D15110300320000) and the International Collaboration Projects (No. 2015DFE12880 and No. B16001).

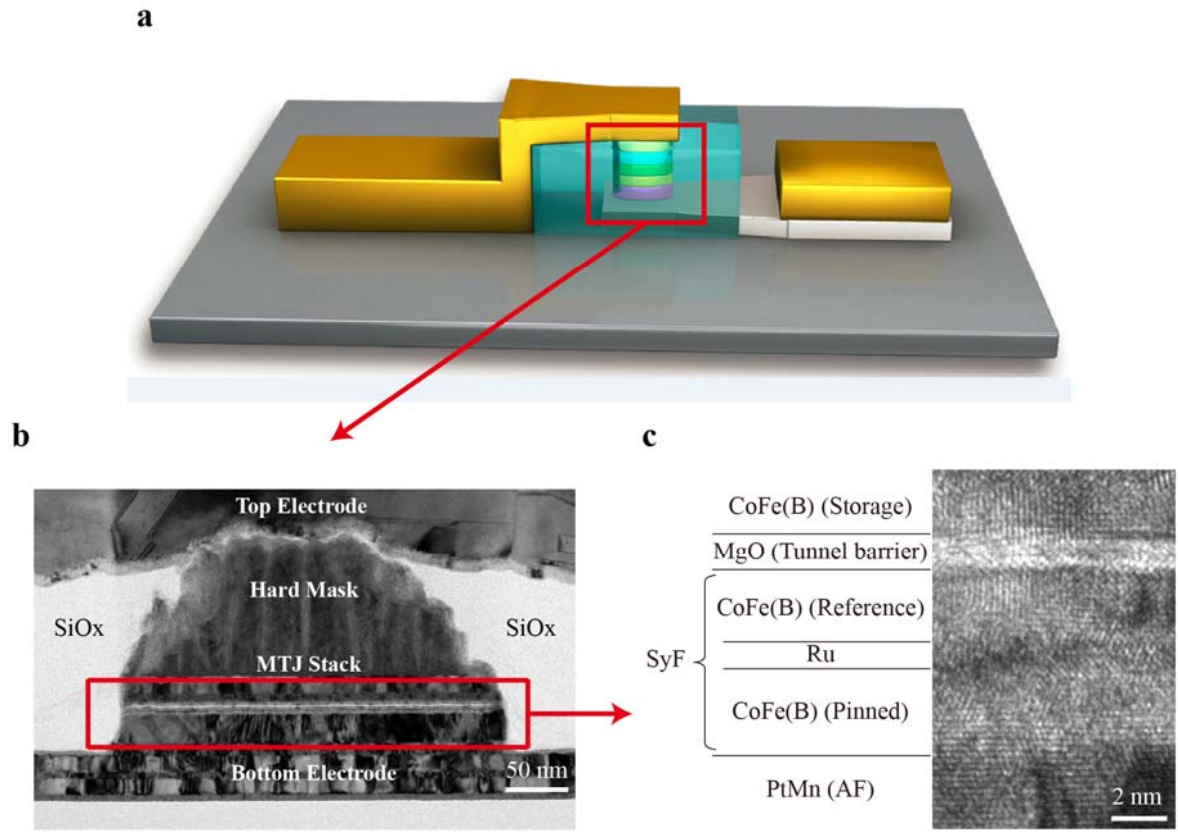

**Figure 1.** Re-MTJ device. (a) Schematic of the Re-MTJ device. A two-terminal MTJ nanopillar is encapsulated within a $SiO_x$-based matrix. (b) TEM image of the Re-MTJ device. The top electrode, hard mask, MTJ stack (indicated with a red rectangular) and bottom electrode are indicated. (c) HRTEM image of a CoFe(B)-MgO-based MTJ nanopillar. The free layer (CoFe(B)), tunnel barrier (MgO), synthetic ferri-magnetic (SyF) reference layer (CoFe(B)/Ru/CoFe(B)) and anti-ferromagnetic layer PtMn are indicated. A crystalline structure can be observed for both CoFe(B) and MgO.

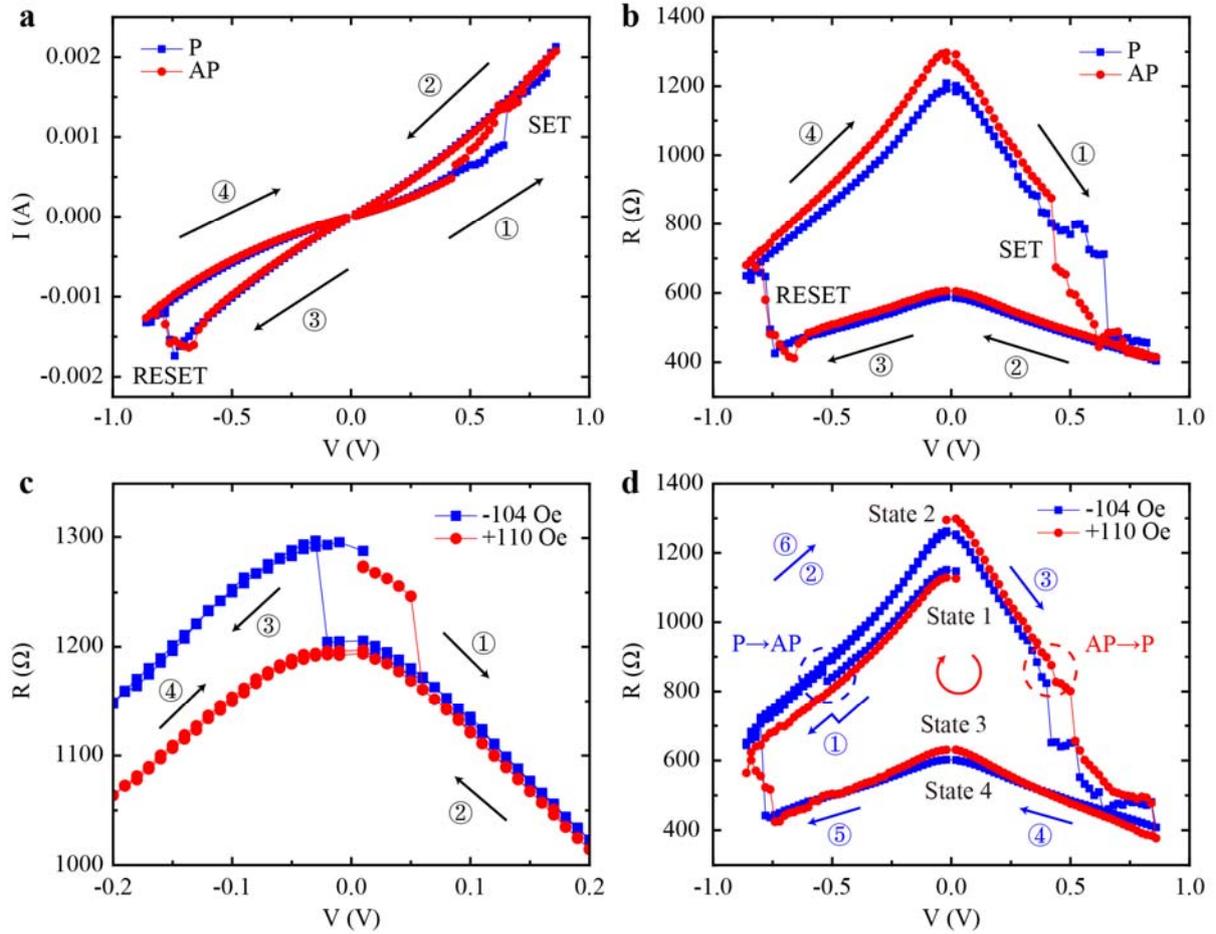

**Figure 2.** Transport properties of the Re-MTJ device. The direction of the external magnetic field is along the easy axis of the ellipse. The arrows and numbers (① ~ ⑥) indicate the voltage sweep direction. States 1 to 4 correspond to the resistance states described in Figure 4b. (a) I-V curves without an applied magnetic field up to ±1 V. The MTJ was set to either the P or AP configuration before the measurement. The SET and RESET voltages are indicated. (b) Corresponding R-V curves of (a). (c) R-V curves under external positive and negative magnetic fields for voltages below ±0.2 V, indicating pure CIMS without VIRS. (d) R-V curves under external positive and negative magnetic fields up to ±1 V. The VIRS process is observed near +0.5 V (SET) and -0.7 V (RESET). For the CIMS process, P to AP switching is observed near -0.5 V for a magnetic field of -104 Oe and near +0.4 V for a magnetic field of +110 Oe.

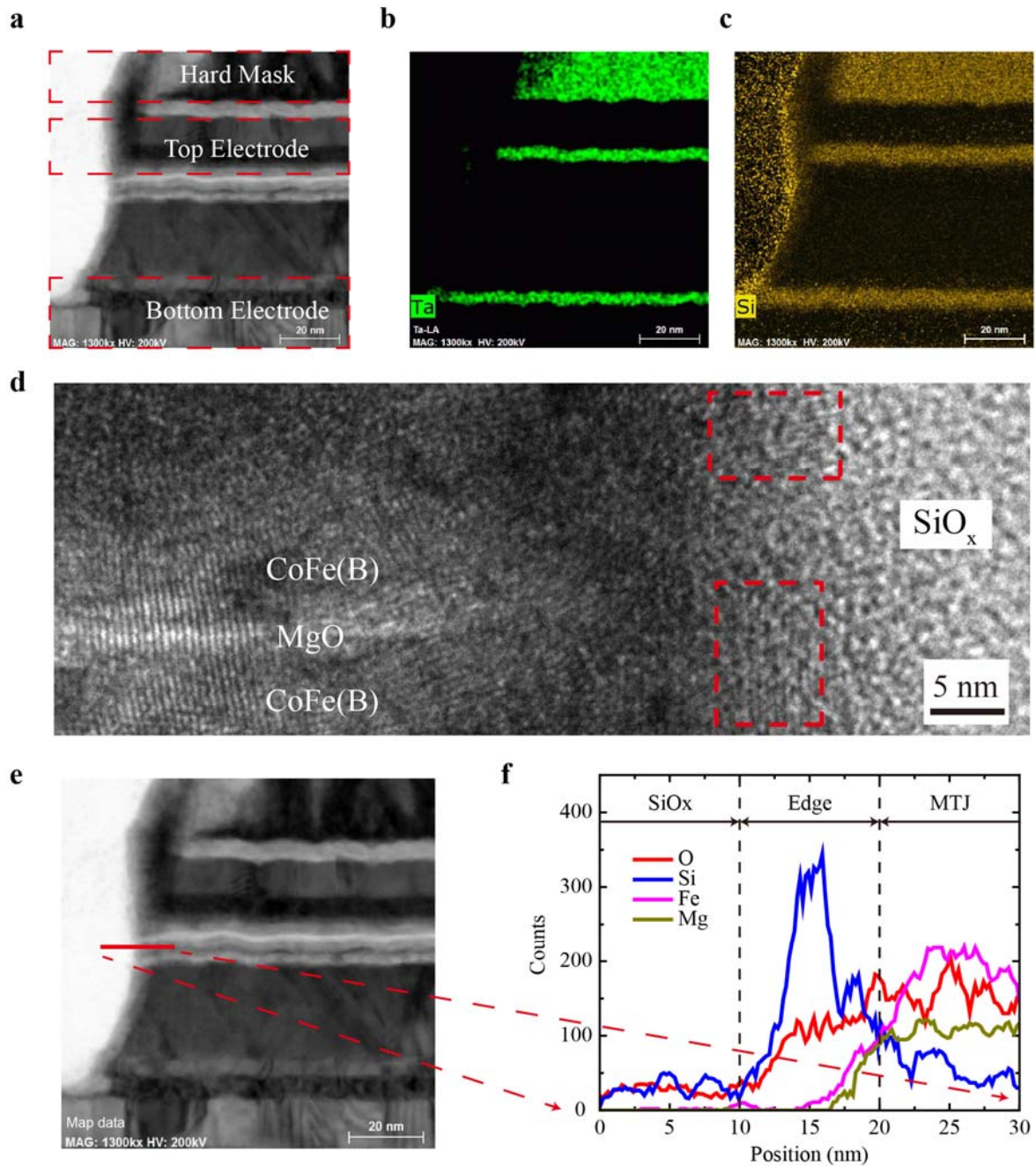

**Figure 3.** Microstructural characterization of the Re-MTJ device. (a) STEM image near the vertical edge of the MTJ nanopillar. The regions of the hard Ta mask, top electrode and bottom electrode are indicated by the red dashed rectangles. (b) EDS mapping of Ta using the $Ta\text{-}L_\alpha$ line characteristic peaks. (c) EDS mapping of Si using the $Si\text{-}K$ series line characteristic peaks. Note that since the $Ta\text{-}M_\alpha$ (1.709 KeV) and $Si\text{-}K_\alpha$ (1.739 KeV) peaks overlap, Ta is also detected. (d) HRTEM image of the edges of the nanopillars. Nanocrystalline structures embedded in the $SiO_x$ matrix are indicated by the red dashed rectangles. (e) STEM image obtained using an HAADF detector. The EDS linescan is marked in red and was measured from the $SiO_x$ matrix into the MTJ nanopillar. (f) EDS linescans for O, Si, Fe and Mg corresponding to the red line indicated in (e). Three different regions can be delimited: a pure $SiO_x$ region, an intermixed layer with aggregates of Si and O (10 nm) and an MTJ region.

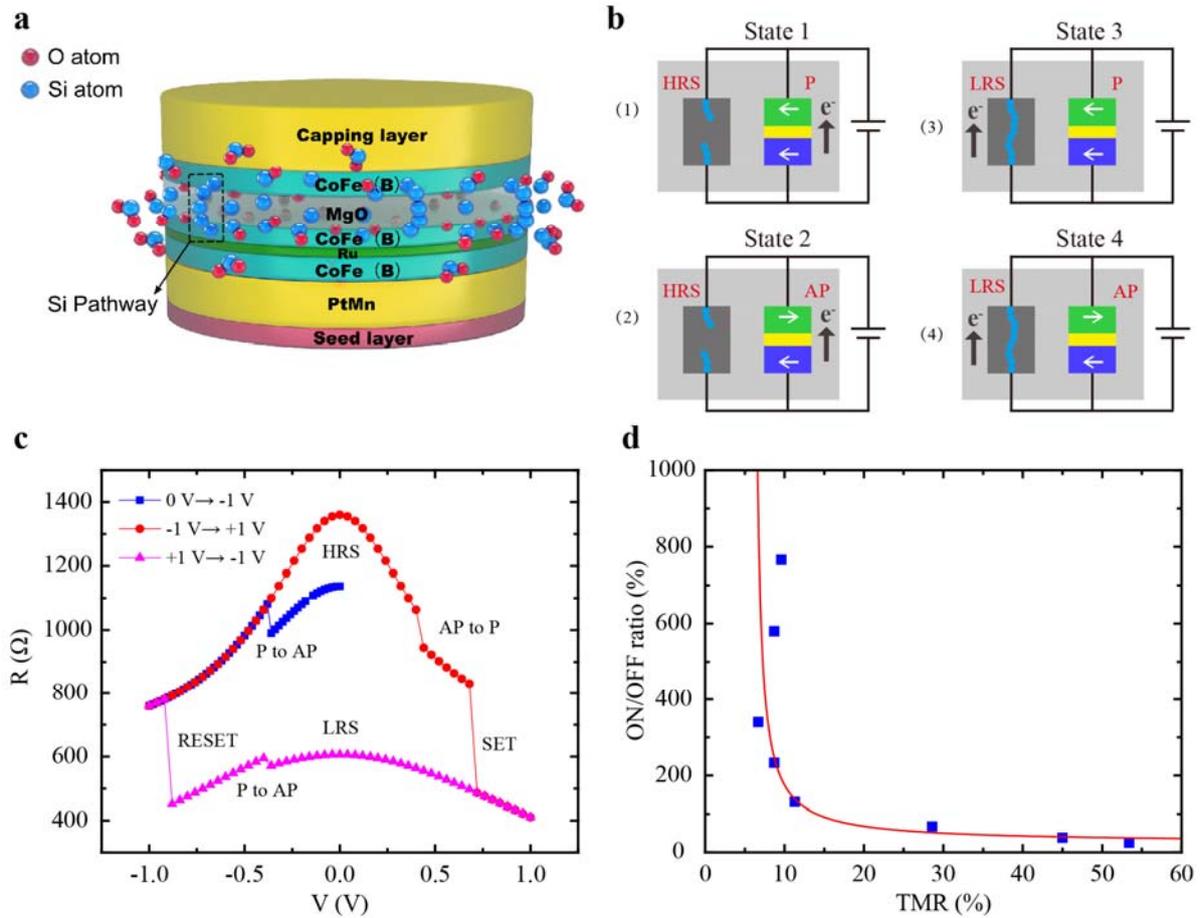

**Figure 4.** Resistance switching model of the Re-MTJ device. (a) Schematic of the MTJ nanopillar surrounded by Si filaments. The blue and red balls represent the Si atoms and O atoms, respectively. (b) Physical model corresponding to an RRAM element in parallel with an MRAM element. Depending on the configuration of the RRAM and MRAM elements, four different states can be obtained in the Re-MTJ device. The blue balls represent the conductive filaments that form the Si pathway. States 1 - 4 correspond to those in Figure 2d. (c) Simulation of the R-V behaviour using a compact model of STT-MTJ and an MIM resistive junction connected in parallel. A magnetic field was not included in the simulation. (d) Relationship between the ON/OFF ratio and the TMR ratio measured from the different devices. The TMR was obtained through R-H measurements under a low voltage of 10 mV. The ON/OFF ratio was obtained after conducting a voltage (current) sweep. The blue squares are the experimental data, and the red line is a guide for the eyes.

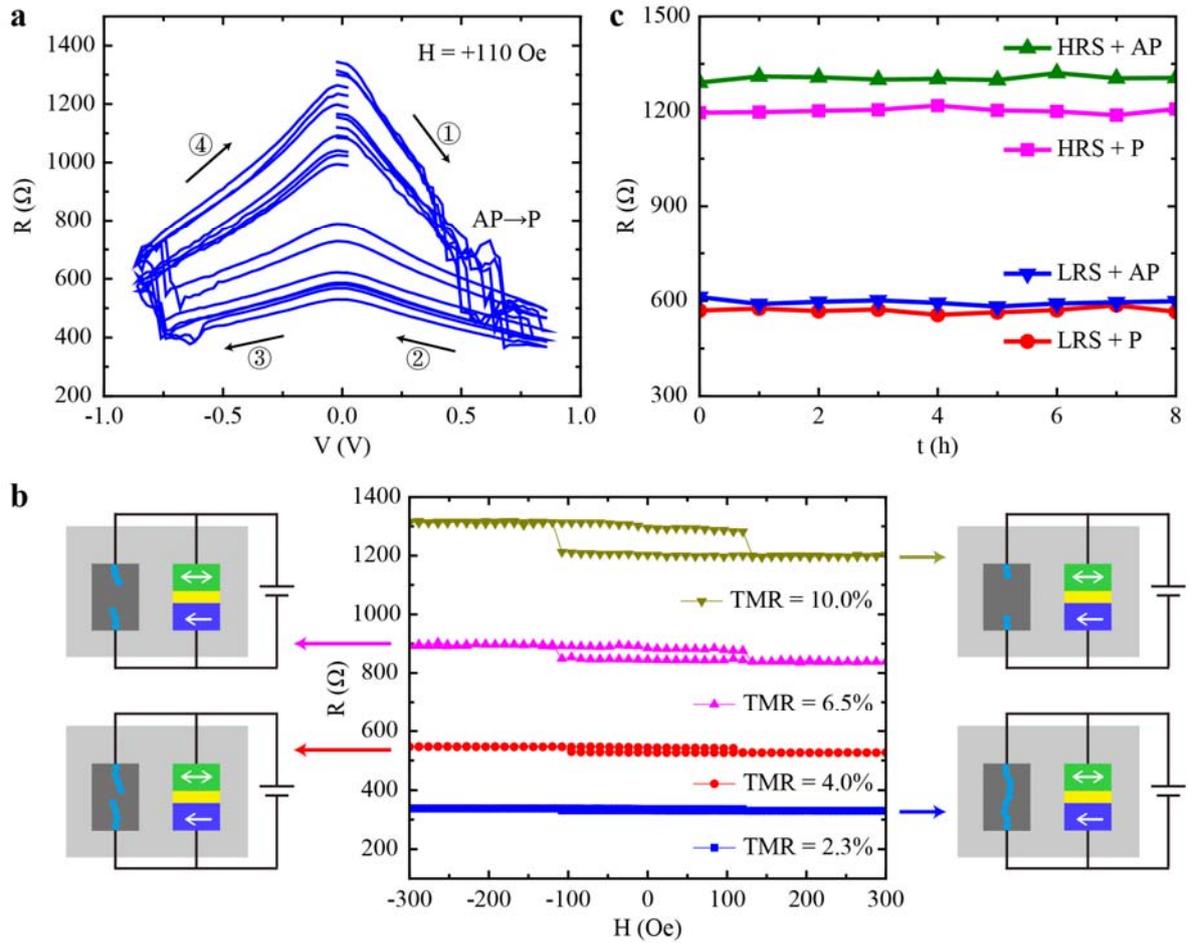

**Figure 5.** Multilevel resistance states in the Re-MTJ device. (a) Seven different R-V curves under external magnetic fields in the same device. (b) R-H hysteresis loops with different initial resistance states for the same device. The measurements were conducted from an LRS (approximately 600 Ω) to an HRS (approximately 1300 Ω). (c) Time-independent resistance curves showing the non-volatile features of the Re-MTJ for four different resistance states (LRS+AP, LRS+P, HRS+AP and HRS+P).

A **nanoscale heterogeneous memristive device** combining the advantages of MRAM and RRAM is demonstrated. The device is based on a resistively enhanced MRAM element integrated with an MTJ nanopillar surrounded by silicon filaments that behave as resistive switches. The device features magnetic switching together with a high ON/OFF ratio of > 1000% and multilevel resistance behaviour.

Keywords: heterogeneous device, resistive switching, magnetic tunnel junction, silicon filaments, spintronics

Yu Zhang, Xiaoyang Lin, Jean-Paul Adam, Guillaume Agnus, Wang Kang, Wenlong Cai, Jean-Rene Coudevylle, Nathalie Isac, Jianlei Yang, Huaiwen Yang, Kaihua Cao, Hushan Cui, Deming Zhang, Youguang Zhang, Chao Zhao, Weisheng Zhao\*, and Dafine Ravelosona

**Heterogeneous Memristive Devices Enabled by Magnetic Tunnel Junction Nanopillars Surrounded by Resistive Silicon Switches**

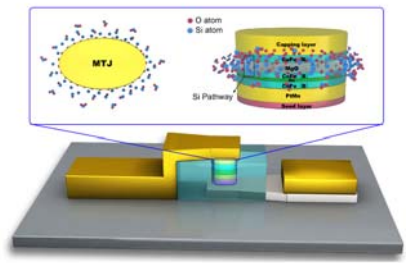

**Supporting Information**

**Heterogeneous Memristive Devices Enabled by Magnetic Tunnel Junction Nanopillars Surrounded by Resistive Silicon Switches**

Yu Zhang, Xiaoyang Lin, Jean-Paul Adam, Guillaume Agnus, Wang Kang, Wenlong Cai, Jean-Rene Coudevylle, Nathalie Isac, Jianlei Yang, Huaiwen Yang, Kaihua Cao, Hushan Cui, Deming Zhang, Youguang Zhang, Chao Zhao, Weisheng Zhao*, and Dafine Ravelosona

**Supporting Note 1: Crystalline structure analysis of the MgO and CoFeB layers of MTJ**

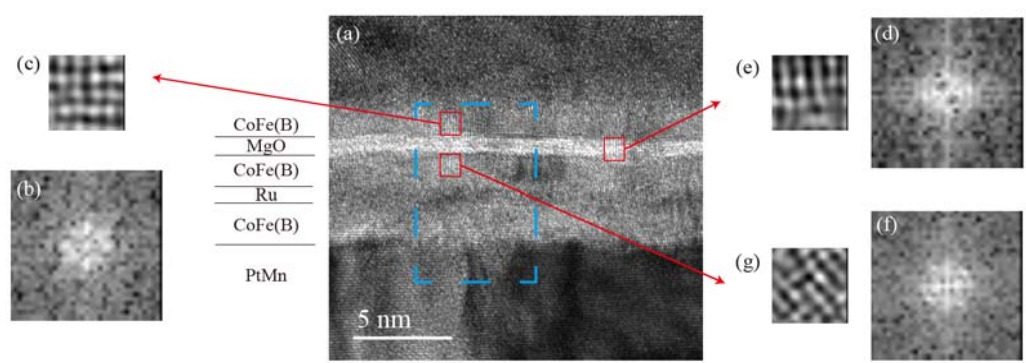

**Figure S1.** The crystalline structure analysis of the MgO and CoFe(B) layers of MTJ. The region in Figure 1c is marked by the blue dash rectangle. The original image (a) were firstly processed by a Gaussian filter (not showed), and then the Fast Fourier transformation (FFT) diffraction patterns (b, d, f) were obtained and finally, the crystalline lattice patterns (c, e, g) after the inversed Fast Fourier transformation.

**Supporting Note 2: ON/OFF ratios in the Re-MTJ devices**

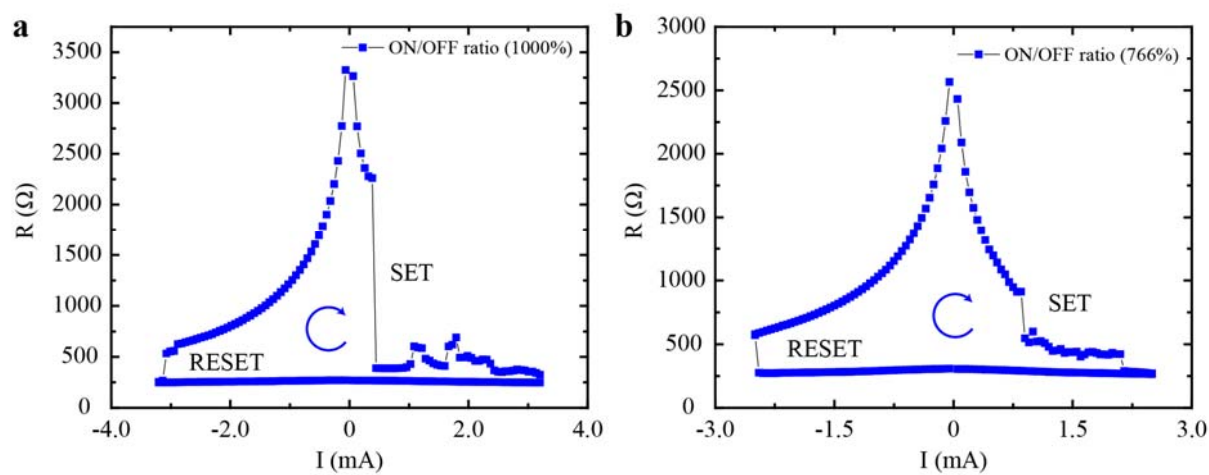

**Figure S2.** Resistance *vs* current in the Re-MTJ devices. High ON/OFF ratios were obtained, such as the two examples shown here that were measured on the same wafer: (a) 1000% and (b) 766%. The ratio most likely depends on the oxidation state of the Si filaments for each device. The arrows indicate the current sweep direction.

**Supporting Note 3: Magnetic properties of the unpatterned magnetic tunnel junction films**

The magnetization curves were measured using a MicroSense vibrating sample magnetometer system at room temperature.

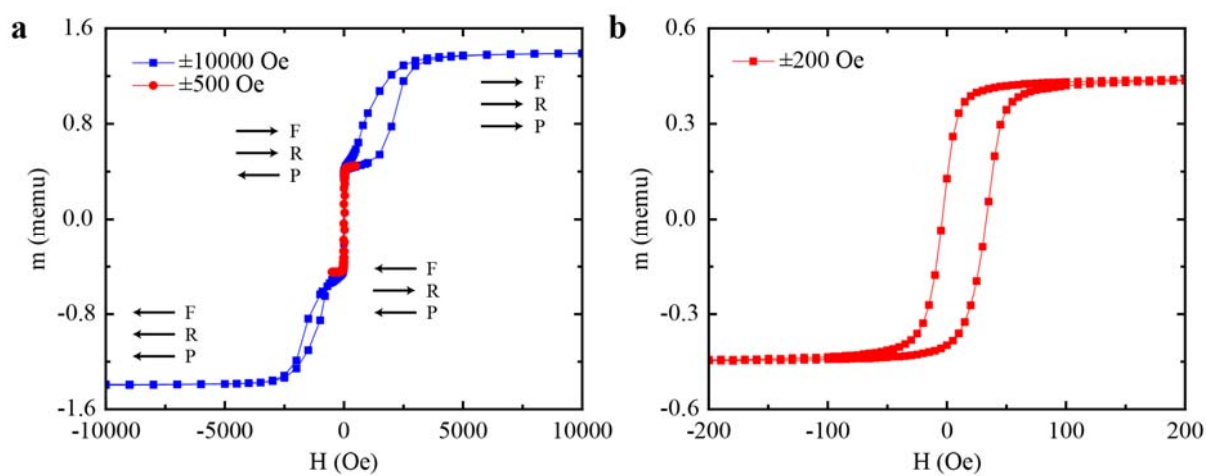

**Figure S3.** In-plane magnetization curves of the unpatterned films. (a) Hysteresis loop for the MTJ structure. The magnetic configurations of the free (F), reference (R) and pinned (P) layers are indicated by arrows. The switching of the free layer is indicated in red. (b) Minor loop corresponding to the switching of the free layer.

**Supporting Note 4: Characterization of O element on the edge of Re-MTJ device**

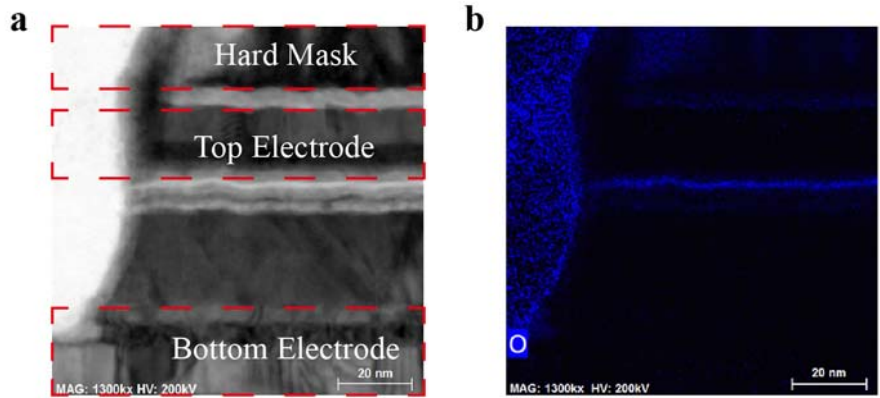

**Figure S4.** (a) HAADF STEM image near the vertical edge of the MTJ nanopillar. The regions of the hard mask, top electrode and bottom electrode are indicated by the red dashed rectangles. (b) EDS mapping of O element according to the same region indicated in (a).

**Supporting Note 5: The resistance gaps between LRS+AP and LRS+P**

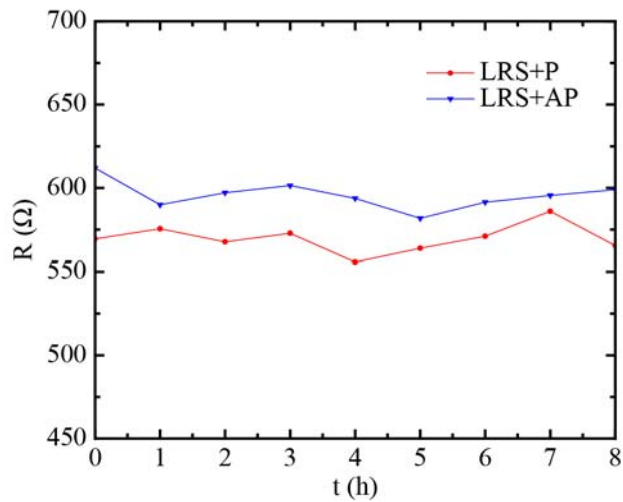

**Figure S5.** Time-independent resistance curves showing the non-volatile features of the Re-MTJ for two resistance states (LRS+AP and LRS+P) when the Si filaments are conductive. The curves are plotted with the same data as Figure 5c.

**Supporting Note 6: Device nanofabrication and sample preparation for TEM**

The Re-MTJ devices were patterned using a self-aligned lift-off and back-end-of-line (BEOL) process. Submicron-sized ellipses were obtained using EBL with a ZEP520A positive resist deposited on top of a 150 nm Ta layer, followed by Pt deposition and a lift-off process. The Pt patterns were used as a protective mask to etch down the 150 nm Ta layer using inductively coupled plasma (ICP). Then, the Ta patterns were used as a hard mask to etch down the MTJs using an optimized IBE process to avoid sidewall redisposition. A VM652 promoter and Accuflo T-25 Spin-on Polymer (produced by Honeywell) were sequentially spin coated, followed by a low-temperature curing process (below 300 °C) for encapsulating the patterned structure in the $SiO_x$. The encapsulation layer was patterned into $40 \times 60$ μm$^2$ elements using ICP. Finally, Cr/Au top electrodes were fabricated utilizing a lift-off approach.

The cross-sectional samples were prepared by using a focused ion beam in the plane of the long axis of the ellipse. The HRTEM, scanning TEM (STEM) and EDS mapping/line scanning were performed using a JEM-ARM-200F transmission electron microscope operating at 200 KeV.